# Performance and Buffering Requirements of Internet Protocols over ATM ABR and UBR Services[1]

Shiv Kalyanaraman, Raj Jain, Sonia Fahmy, Rohit Goyal
The Ohio State University
Department of CIS
Columbus, OH 43210-1277
Email: {*shivkuma, jain, fahmy, goyal*}@cis.ohio-state.edu
and
Seong-Cheol Kim
Principal Engineer, Network Research Group
Communication Systems R&D Center
Samsung Electronics Co. Ltd.
Chung-Ang Newspaper Bldg.
8-2, Karak-Dong, Songpa-Ku
Seoul, Korea 138-160
Email: kimsc@metro.telecom.samsung.co.kr,

## Abstract

The Asynchronous Transfer Mode (ATM) networks are quickly being adopted as backbones over various parts of the Internet. This paper analyzes the performance of TCP/IP protocols over ATM network's Available Bit Rate (ABR) and Unspecified Bit Rate (UBR) services. It is shown that ABR pushes congestion to the edges of the ATM network while UBR leaves it inside the ATM portion.

# 1 Introduction

With the proliference of multimedia traffic over the Internet, it seems natural to move over to ATM technology which has been designed specifically to support integration of data, voice, and video applications. While multimedia applications are still in the development stage, most of the traffic on the Internet today is data traffic in the sense that they are bursty and relatively delay insensitive. It is, therefore, natural to ask how the current applications will perform over ATM technology.

Although ATM technology has been designed to provide an end-to-end transport level service and so, strictly speaking, there is no need to have TCP or IP if the entire path from source to destination is an ATM path. However, in the forseeable future, this scenario is going to be rare. A more common scenario would be where only part of the path is ATM.

In this case, TCP is needed to provide the end-to-end transport functions (like flow control, retransmission, ordered delivery) and ATM networks are used simply as "bit pipes" or "bitways."

ATM networks provide multiple classes of service. Of these, the Available Bit Rate (ABR) and the Unspecified Bit Rate (UBR) service classes have been developed specifically to support data applications. The ABR service requires network switches to constantly monitor their load and feed the information back to the sources, which in turn dynamically adjust their input into the network. For UBR service, the switches monitor their queues and simply discard cells or packets of overloading users. Intelligent users may use this packet loss as an implicit feedback indicating network congestion and reduce their input to the network. The Transport Control Protocol (TCP) does have this intelligence built into it in the form of "Slow Start" congestion avoidance mechanism [7]. Therefore, there is currently a debate in the networking community about the need for ABR service particularly in light of TCP's built-in congestion control facilities.

In the Internet, at least The choice of ABR vs UBR is that of intelligent bit pipe vs intelligent transport. With ABR, ATM networks control the congestion intelligently and fast. With UBR ATM switches behave similar to legacy routers and most of the congestion control is exercised by TCP. It is interesting to compare the performance of TCP over ABR and UBR. We find that TCP performs best when it does not experience packet loss. We quantify the amount of buffering required at the ATM switches to avoid TCP packet loss. Notably, ABR is scalable over TCP in the sense that it requires buffering which does not depend upon the number of connections. The amount of buffering depends upon factors such as the switch congestion control scheme used, and the maximum round trip time (RTT) of all virtual circuits (VCs) through the link. UBR is not scalable in the sense that it requires buffering proportional to the sum of the TCP receiver windows of all sources.

## 2  ABR Traffic Management: An Overview

The ATM Forum completed the Traffic Management Version 4.0 (TM4.0) specification in April 1996 [6]. The document specifies five classes of service: constant bit rate (CBR), real-time variable bit rate (rt-VBR), non-real time variable bit rate (nrt-VBR), available bit rate (ABR), and unspecified bit rate (UBR). The CBR and VBR services provide quality of service(QoS) guarantees to support delay-sensitive applications such as voice, video, and multimedia applications. The ABR and UBR services provide efficient sharing of the remaining bandwidth to support non-delay sensitive data applications. Link bandwidth is first allocated to the VBR and CBR classes. The remaining bandwidth, if any, is given to ABR and UBR traffic.

The components of the ABR traffic management framework are shown in Figure 1. In order to obtain the network feedback, the sources send resource management (RM) cells after every $n$ (default 31) data cells. The destination simply returns these RM cells back to the source. The RM cells contain an explicit rate (ER) field in which switches along the path to indicate the rate that the source should use after the receipt of the RM cell.

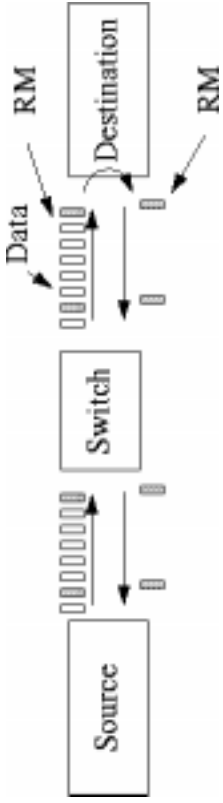

Figure 1: ABR Traffic Management Model: Source, Switch, Destination and Resource Management Cells

One advantage of this explicit rate feedback is that each switch can calculate the rate which it wants to allocate to the flow by its own method and reduce the ER field if necessary. The switches are not allowed to increase the field. There is no need to standardize a particular switch algorithm and TM4.0 specification does not specify any standard switch algorithm.

While not requiring a switch algorithm is a feature of explicit rate feedback mechanism, the peformance of the mechanism is dependent heavily on the switch algorithm. Some switch algorithms are very slow to respond to traffic changes while others may be too fast. Some switch algorithms are unfair under certain circumstances while others would be fair under those circumstances.

Our study uses the ERICA switch algorithm, which is included in TM4.0 as an example of possible switch algorithms. All statements about the performance, scalability, and buffering requirements of ABR service, in this paper, are based on the use of ERICA switch algorithm and its variants as described in [3].

TCP is the most popular transport protocol for data transfer. It provides a reliable transfer of data using a window-based flow and error control algorithm [7]. TCP runs over IP which in turn can run over ATM. When TCP uses the ABR service, there are two control algorithms active: the TCP window-based control running on top of the ABR rate-based control. On the other hand, when TCP uses the UBR service, only the TCP flow control is active. It is important to verify that TCP/IP performs satisfactorily over ABR and UBR.

In our experiments, we use ftp-like (unidirectional and large data transfer) applications running over TCP. We initially study the conditions under which TCP can achieve maximum throughput. We find that TCP performs best when there is no packet loss, and sufficient buffering at the switches is the key requirement for both ABR and UBR services. Next, we quantify the buffering requirement and study the factors affecting it. We find that the ABR service is scalable in terms of number of sources (or virtual circuits). The total buffer required in a switch to achieve zero loss is bounded. This bound depends upon the RTT of VCs but not on their number. The UBR service is not scalable in terms of the number of sources, though the buffering requirement is bounded by the sum of the TCP maximum window sizes.

While this paper studies TCP under zero packet loss conditions, other papers [1, 8, 9] suggest methods to improve TCP performance over ATM under lossy conditions.

# 3 TCP Congestion Mechanisms

TCP is one of the few transport protocols that has its own congestion control mechanisms. The key TCP congestion mechanism is the so called "Slow start" [7]. TCP connections use an end-to-end flow control window to limit the number of packets that the source sends. The sender window is the minimum of the receiver window (Wrcvr) and a congestion window variable (CWND).

Whenever a TCP connection loses a packet, the source does not receive an acknowledgment (ack) and it times out. The source remembers the congestion window (CWND) value at which it lost the packet by setting a threshold variable SSTHRESH at half the window. More precisely, SSTHRESH is set to max{2, min{CWND/2, Wrcvr}} and CWND is set to one.

The source then retransmits the lost packet and increases its congestion window by one every time a packet is acknowledged. We call this phase the "exponential increase phase" since the window when plotted as a function of time increases exponentially. This continues until the window is equal to SSTHRESH. After that, the window $w$ is increased by $1/w$ for every packet that is acked. This is called the "linear increase phase" since the window graph as a function of time is approximately a straight line. Note that although the congestion window may increase beyond the advertised receiver window, the source window is limited by that value. when packet losses occur, the retransmission algorithm may retransmit all the packets starting from the lost packet. That is, TCP uses a go-back-N retransmission policy. The typical changes in the source window plotted against time are shown in Figure 2.

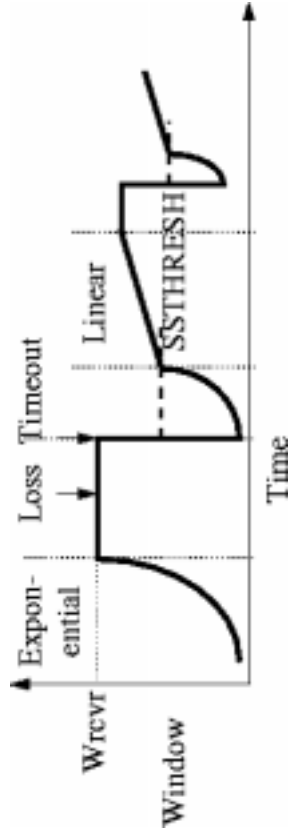

Figure 2: TCP Window vs Time using Slow Start

When there is a bursty loss due to congestion, time is lost due to timeouts and the receiver may receive duplicate packets as a result of the go-back-N retransmission strategy. This is illustrated in Figure 3. Packets 1 and 2 are lost but packets 3 and 4 make it to the destination are are stored there. After the timeout, the source sets its window to 1 and retransmits packet 1. When that packet is acknowledged, the source increases its window to 2 and sends packets 2 and 3. As soon as the destination receives packet 2, it delivers all packets upto 4 to the application and sends an ack (asking for packet 5) to the source. The 2nd copy of packet 3, which arrives a bit later is discarded at the destination since it is a duplicate.

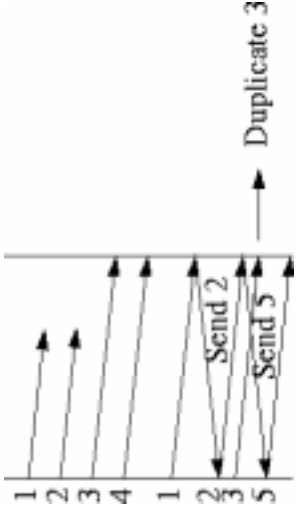

Figure 3: Timeout and Duplicate Packets in Slow Start

## 4 Behavior of TCP over UBR

The UBR service class does not include flow control and hence depends upon transport layers to provide flow control. When TCP uses UBR, and cells are dropped at the ATM layer, TCP has to recover from the resulting packet drops using its congestion mechanisms.

When the ATM switch has limited buffers, a single cell drop at the ATM level results in a packet drop in the TCP level [9]. This phenomenon can result in low throughput and unfairness for TCP connections. When a cell is dropped, the destination drops an entire packet. TCP then times out and retransmits the entire packet. The low TCP throughput is due to the time lost in the timeouts and retransmits of various packets.

Maximum TCP thoughput over UBR is observed when the switches have sufficient buffering such that TCP does not lose packets. However, even with limited buffering, TCP throughput and fairness over UBR can be improved by proper buffer allocation, drop policies and scheduling:

- **Drop policies** decide when to drop cells. Drop policies are critical in UBR to achieve good throughput. The early packet discard (EPD) [9] policy drops full packets instead of random cells from multiple packets.

- **Buffer allocation** policies decide how to divide the available buffers among the cells from contending connections. Buffer allocation can be sophisticated even though the buffer service policy is simple, for example, first in first out (FIFO). Fair buffer allocation (FBA) schemes improve fairness by selectively discarding frames from flows that are sending more than their fair share.

- **Scheduling policies** divide the available bandwidth among various contending classes. Scheduling may be implemented at a coarse granularity (per-class scheduling) to divide bandwidth among the different service classes (CBR, rt-VBR, nrt-VBR, ABR, and UBR) or at a fine level granularity (per-VC scheduling) to divide bandwidth between various connections in a service class. The per-class scheduling decides how much bandwidth UBR connections get, after higher priority classes are served. The per-VC scheduling can control the bandwidth distribution among contending UBR connections. Specifically, cells from a "rogue" connection sending more than a fair share of bandwidth receive only their fair share and may be delayed more than the cells of other connections.

# 5  Buffering Requirements for TCP over UBR

UBR provides no flow control of its sources. Hence, when the switches are overloaded, the network queues will build up unless the transport layer controls its transmission. A TCP source stops increasing its transmission rate only when its congestion window reaches a maximum value. This maximum window size is characterized by the parameter MAXWIN. The default maximum congestion window, MAXWIN, is 65536 bytes (64 kB). However, in high delay links, this window is too small to achieve full throughput. On the other hand, having maximum window above the round trip time (RTT) is not useful. This is because at any time, only 1 RTT worth of segments can be in the TCP pipe. The maximum window size determines the amount of data that can be present in the TCP pipe. As a result, this parameter determines the storage capacity needed in the network.

Specifically, TCP using UBR requires network buffers equal to the sum of the maximum window sizes of all the TCP connections to avoid cell loss. In this respect, UBR is not scalable for TCP. Note that this result is unaffected by other factors such as the round trip time and the topology configuration of the network.

# 6  Behavior of TCP over ABR

## 6.1  Closed Loop vs Open Loop Control

In contrast to UBR, the ABR service provides flow control at the ATM level itself. Based on their load, ABR switches return feedback in RM cells. The RM cells return to the source carrying the minimum of the feedback indications from all the switches. When the source receives the RM cell (feedback) from the network, it adjusts its rate according to the feedback. When there is a steady flow of RM cells in the forward and reverse directions, there is a steady flow of feedback from the network. In this state, we say that the ABR control loop has been established and the source rates are primarily controlled by the network feedback (closed-loop control). The network feedback is effective after a time delay. The time delay required for the new feedback to take effect is the sum of the time taken for an RM cell to reach the source from the switch and the time for a cell (sent at the new rate) to reach the switch from the source. This time delay is called the "feedback delay."

When the source transmits data after an idle period, there is no reliable feedback from the network. For one round trip time (time taken by a cell to travel from the source to the destination and back), the source rates are primarily controlled by the ABR source end system rules (open-loop control). The open-loop control is replaced by the closed-loop control once the control loop is established. When the traffic on ABR is "bursty" i.e., the traffic consists of busy and idle periods, open-loop control may be exercised at the beginning of every active period (burst). Hence, the source rules assume considerable importance in ABR flow control.

## 6.2 Nature of TCP Traffic at the ATM Layer

Data which uses TCP is controlled first by the TCP "slow start" procedure before it appears as traffic to the ATM layer. Suppose we have a large file transfer running on top of TCP. When the file transfer begins, TCP sets its congestion window (CWND) to one. The congestion window increases exponentially with time. Specifically, the window increases by one for every ack received. Over any round trip time (RTT), the congestion window doubles in size.

As shown in Figure 4, at the ATM layer, the TCP traffic is considered bursty. Initially, there is a short active period (the first packet is sent) followed by a long idle period (nearly one round-trip time, waiting for an ACK). The length of the active period doubles every round-trip time and the idle period reduces correspondingly. Finally, the active period occupies the entire round-trip time and there is no idle period. After this point, the TCP traffic appears as an infinite (or persistant) traffic stream at the ATM layer.

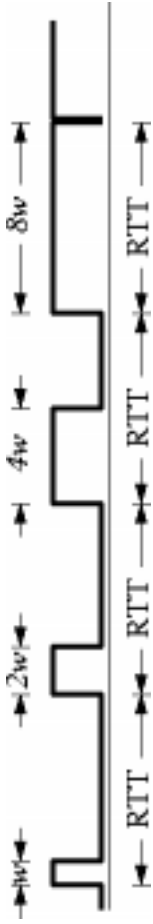

Figure 4: At the ATM layer, the TCP traffic results in bursts. The burst size doubles every round trip until the traffic becomes continuous.

When sufficient load is not experienced at the ABR switches, the switch algorithms typically allocate high rates to the sources. This is likely to be the case when a new TCP connection starts sending data. The file transfer data is bottlenecked by the TCP congestion window size and not by the ABR source rate. In this state, we say that the TCP sources are *window-limited*.

The TCP active periods double every round trip time and appear as infinite traffic at the ATM layer. The TCP congestion window is now large and is increasing. Hence, it will send data at rate greater than the source's sending rate. The file transfer data is bottlenecked by the ABR source rate and not by the TCP congestion window size. In this state, we say that the TCP sources are *rate-limited*.

The ABR queues at the switches start increasing when the TCP idle times are not sufficient to clear the queues built up during the TCP active times. The queues may increase until the ABR source rates converge to optimum values. Once the TCP sources are rate-limited and the rates converge to optimum values, the lengths of the ABR queues at the switch will start decreasing. TCP achieves maximum throughput over ABR when there is no cell loss.

## 6.3 TCP Performance With Cell Loss

Cell loss will occur in the network if the ATM switches do not have sufficient buffers to accomodate this queue buildup. In a detailed study, we find that TCP achieves maximum throughput over ABR when there is no cell loss [2]. When cell loss does occur, the cell loss ratio (CLR) metric, which quantifies cell loss, is a poor indicator of loss in TCP throughput. This is because TCP loses time (through timeouts) rather than cells (cell loss). Smaller TCP timer granularity (which controls timeout durations) can help improve throughput. If the ABR rates do not converge to optimum values before the cell loss occurs, the effect of the switch congestion scheme may be dominated by factors such as the timer granularity. Intelligent drop policies can help improve the throughput slightly.

TCP Reno, a widespread version of TCP, includes the Fast Retransmit and Fast Recovery algorithms that improve TCP performance when a single segment is lost. However, in high bandwidth links, network congestion results in several dropped segments. In this case, fast retransmit and recovery are not able to recover from the loss and slow start is triggered. Fast retransmit and recovery are effective in single packet losses typically due to error. In our experiments, all losses are due to congestion and result in multiple segments being dropped.

The TCP throughput loss over ABR can be avoided by provisioning sufficient switch buffers. We show in a later section that the buffer requirement is bounded and small. However, note that, even after ABR sources converge to optimum rates, the TCP congestion window can grow till it reaches its maximum (negotiated) value. In such cases, TCP overloads the ABR source and the queues build up at the source end system. If the source queues overflow cell loss will occur, and performance will degrade. In this case, the cell loss occurs outside the ABR network.

## 7  Buffering Requirements for TCP over ABR

Using simulations we found that the maximum buffer requirement for TCP over ABR without any loss is approximately $(a \times \text{RTT} + c \times \text{feedback delay}) \times \text{link bandwidth}$ for low values of coefficients $a$ and $c$. In fact, for most cases with ERICA algorithm, we found the coefficient $a$ to be 3. This can be easily justified as follows:

- Initially the TCP load doubles every RTT. During this phase, TCP sources are window-limited [2], i.e., their data transmission is bottlenecked by their congestion window sizes and not by the network directed rate.

- The minimum number of RTTs required to reach rate-limited operation [2] decreases as the logarithm of the number of sources. In other words, the more the number of sources, the faster they all reach rate-limited operation. Rate-limited operation occurs when the TCP sources are constrained by the network directed ABR rate rather than their congestion window sizes.

- After the pipe just becomes full (TCP keeps sending data for one RTT), the maximum queue which can build up before fresh feedback reaches the sources is $1 \times$ RTT $\times$ link bandwidth. This observation follows because the aggregate TCP load can at most double every RTT and fresh feedback reaches sources every RTT.

- Queue backlogs due to TCP bursts smaller than RTT (before the pipe became full) are $1 \times$ RTT $\times$ link bandwidth. The TCP idle periods are not sufficient to drain out the queues built up during the TCP active periods. This occurs when the idle periods are shorter than the active periods. Given that TCP load doubles every RTT, the backlog is at most $1 \times$ RTT $\times$ link bandwidth.

- When two-way traffic is used, TCP behaves in a more bursty fashion. This is because the acks may be received in bursts, rather than being spaced out over time. This bursty behavior of acks causes an additional $1 \times$ RTT $\times$ link bandwidth queues. When acks are bursty, the doubling of the TCP load can occur instantaneously (not spaced over time) and an extra round-trip worth of queues are built up.

- Once load is experienced continuously at the switch, the TCP sources appear as infinite sources to the switch. The switch algorithm then takes $c$ feedback delays to converge to the max-min rates (when the queue length is guaranteed to decrease). Assuming that the TCP sources are rate-constrained during the convergence period, the aggregate TCP load can only decrease. In the worst case, the queue built up during the convergence phase is $c \times$ feedback delay $\times$ link bandwidth.

Since feedback delay is at most RTT, the sum of these components is approximately ($3 \times$ RTT $+ c \times$ feedback delay) $\times$ link bandwidth.

Though the maximum ABR network queues are small, the queues at the sources are high. Specifically, the maximum sum of the queues in the source and the switches is equal to the sum of the TCP window sizes of all TCP connections. In other words the buffering requirement for ABR becomes the same as that for UBR if we consider the source queues into consideration. This observation is true only in certain ABR networks. If the ATM ABR network is an end-to-end network, the source end systems can directly flow control the TCP sources. In such a case, the TCP will do a blocking send, i.e., and the data will go out of the TCP machine's local disk to the ABR source's buffers only when there is sufficient space in the buffers.

The ABR service may also be offered at the backbone networks, i.e., between two routers. In these cases, the ABR source cannot directly flow control the TCP sources. The ABR flow control moves the queues from the network to the sources. If the queues overflow at the source, TCP throughput will degrade.

# 8 Factors Affecting ABR Buffering Requirements

In this section we present sample simulation results to substantiate the preceding claims and analyses. We also analyze the effect of some important factors affecting the ABR buffering requirements. The key metric we observe is the maximum queue length.

## 8.1 The $n$ Source Configuration

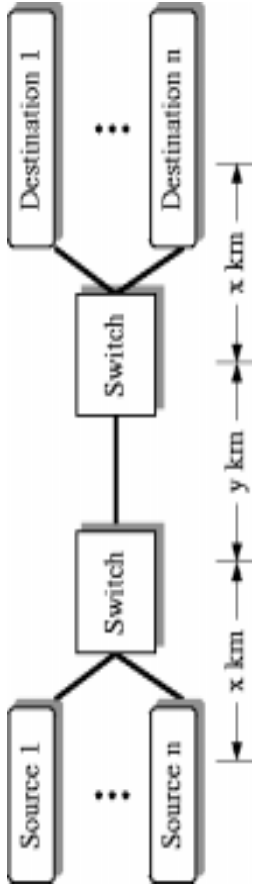

Figure 5: $n$ Source Configuration

All our simulations presented use the $n$ Source configuration. The $n$ Source configuration has a single bottleneck link shared by $n$ ABR sources. All links run at 155.52 Mbps and are of the same length. We experiment with the number of sources and the link lengths.

All traffic is unidirectional. A large (infinite) file transfer application runs on top of TCP for the TCP sources. N may assume values 1, 2, 5, 10, 15 and the link lengths 1000, 500, 200, 50 km. The maximum queue bounds also apply to configurations with heterogenous link lengths.

## 8.2 TCP and ABR Options

Our experiments use an infinite TCP source running on TCP over ATM. The TCP source always has a frame to send. However, due to TCP window constraint, the resulting traffic at the ATM layer may or may not be continuous. We use a TCP maximum segment size (MSS) of 512 bytes. The window scaling option is used so that the throughput is not limited by path length. The TCP window is set at $16 \times 64$ kB = 1024 kB. For satellite round trip (550 ms) simulations, the window is set using the TCP window scaling option to $34000 \times 2^8$ bytes.

The results presented here use the ERICA algorithm at the ATM switch [3]. The ERICA algorithm uses two key parameters: *target utilization* and *averaging interval length*. The algorithm measures the load and number of active sources over successive averaging intervals and tries to achieve a link utilization equal to the target. The averaging intervals end either after the specified length or after a specified number of cells have been received, whichever happens first. In the simulations reported here, the target utilization is set at 90%, and the averaging interval length defaults to 100 ABR input cells or 1 ms, represented as the tuple (1 ms,100 cells). Default values are chosen for the ABR source parameters [6].

## 8.3 Effect of Number of Sources

In Table 1, we notice that although the buffering required increases as the number of sources is increased, the amount of increase slowly decreases. As later results will show, three RTTs worth of buffers are sufficient even for large number of sources. In fact, one RTT worth of buffering is sufficient for many cases: for example, the cases where the number of sources is small. The rate allocations among contending sources were found to be fair in all cases.

Table 1: Effect of number of sources

| Number of Sources | RTT(ms) | Feedback delay(ms) | Max Q (cells) | Thoughput |
|---:|---:|---:|---:|---:|
| 5 | 30 | 10 | 10597 = 0.95*RTT | 104.89 |
| 10 | 30 | 10 | 14460 = 1.31*RTT | 105.84 |
| 15 | 30 | 10 | 15073 = 1.36*RTT | 107.13 |

## 8.4 Effect of Round Trip Time (RTT)

From Table 2, we find that the maximum queue approaches $3 \times$ RTT $\times$ link bandwidth, particularly for metropolitan area networks (MANs) with RTTs in the range of 6 ms to 1.5 ms. This is because the RTT values are lower and in such cases, the effect of switch parameters on the maximum queue increases. In particular, the ERICA averaging interval parameter is comparable to the feedback delay.

Table 2: Effect of Round Trip Time (RTT)

| Number of Sources | RTT(ms) | Feedback Delay (ms) | Max Q size(cells) | Thoughput |
|---:|---:|---:|---:|---:|
| 15 | 30 | 10 | 15073 = 1.36*RTT | 107.13 |
| 15 | 15 | 5 | 12008 = 2.18*RTT | 108.00 |
| 15 | 6 | 2 | 6223 = 2.82*RTT | 109.99 |
| 15 | 1.5 | 0.5 | 1596 = 2.89*RTT | 110.56 |

## 8.5 LANs: Effect of Switch Parameters

In Table 3, the number of sources is kept fixed at 15. The averaging interval is specified as a pair (T, n), where the interval ends when either T ms have expired or N cells have been processed, whichever happens first. For the parameter values shown in the table, the number of cells determined the length of the averaging interval since under continuous traffic 1000 ATM cells take only 2.7 ms.

Table 3: Effect of Switch Parameter (Averaging Interval)

| Averaging Interval (ms,cells) | RTT(ms) | Feedback Delay (ms) | Max Q size(cells) | Thoughput |
|---|---|---|---|---|
| (10,500) | 1.5 | 0.5 | 2511 | 109.46 |
| (10,1000) | 1.5 | 0.5 | 2891 | 109.23 |
| (10,500) | 0.030 | 0.010 | 2253 | 109.34 |
| (10,1000) | 0.030 | 0.010 | 3597 | 109.81 |

From Table 3, we observe that the effect of the switch parameter, averaging interval, dominates in LAN configurations. The ERICA averaging interval is much greater than the RTT and feedback delay and it determines the congestion response time and hence the queue lengths. configurations. The ERICA averaging interval becomes much greater than

## 8.6 Effect of Feedback Delay

We conducted a 3 × 3 full factorial experimental design to understand the effect of RTT and feedback delays [5]. The results are summarized in Table 4. The thoughput figures for the last three rows (550 ms RTT) are not available since the throughput did not reach a steady state although the queues had stabilized.

Observe that the queues are small when the feedback delay is small and do not increase substantially with round-trip time. This is because the switch scheme limits the rate of the sources before they can overload for a substantial duration of time.

## 9 Summary

The main results of the study are:

1. TCP achieves maximum throughput over ABR and UBR when there are enough buffers at the switches and no cells are lost in the ATM network.

Table 4: Effect of Feedback Delay

| RTT(ms) | Feedback Delay (ms) | Max Q size(cells) | Thoughput |
|---|---|---|---|
| 15 | 0.01 | 709 | 113.37 |
| 15 | 1 | 3193 | 112.87 |
| 15 | 10 | 17833 | 109.86 |
| 30 | 0.01 | 719 | 105.94 |
| 30 | 1 | 2928 | 106.9 |
| 30 | 10 | 15073 | 107.13 |
| 550 | 0.01 | 2059 | NA |
| 550 | 1 | 15307 | NA |
| 550 | 10 | 17309 | NA |

2. When maximum throughput is achieved, the TCP sources are rate-limited by ABR rather than window-limited by TCP.

3. When the number of buffers is smaller, there is a large reduction in throughput even though CLR is very small.

4. The reduction in throughput is due to loss of time during timeouts (large timer granularity), and transmission of duplicate packets which are dropped at the destination.

5. ABR service is scalable in terms of the number of TCP/IP sources. ABR switches with buffers equal to a small multiple of network diameter can guarantee no loss even for a very large number of VCs carrying TCP/IP traffic.

6. UBR service is *not* scalable in terms of the number of TCP/IP sources (or VCs). UBR switches require buffers proportional to the sum of the TCP receiver window sizes.

7. For ABR, the choice of an switch scheme, and providing sufficient buffering is more critical than the choice of a drop policy.

8. The ABR source queues may be high in backbone ABR networks.

---

[2] All our papers and ATM Forum contributions are available through http://www.cis.ohio-state.edu/~jain